\documentclass{aa}  

\usepackage{graphicx}
\usepackage{txfonts}
\usepackage{hyperref}
\begin{document}

   \title{Non-local thermal transport impact on compressive waves in two-temperature coronal loops}

   \subtitle{}

   \author{
            S. A. Belov\inst{\ref{aff:Warwick}}\thanks{Sergey.Belov@warwick.ac.uk},
            T. Goffrey\inst{\ref{aff:Warwick},\ref{aff:HTI}},
            T. D. Arber\inst{\ref{aff:Warwick}},
            \and
            D. Y. Kolotkov\inst{\ref{aff:Warwick},\ref{aff:VIRAC}}  
          }

   \institute{
   \label{aff:Warwick}{Centre for Fusion, Space  and Astrophysics, Department of Physics, University of Warwick, Coventry CV4 7AL, UK}\and
   \label{aff:HTI}{Healthcare Technology Institute, School of Chemical Engineering, The University of Birmingham, UK}\and
    \label{aff:VIRAC}{Engineering Research Institute \lq\lq Ventspils International Radio Astronomy Centre (VIRAC)\rq\rq, Ventspils University of Applied Sciences, Ventspils, LV-3601, Latvia}
    }

   \date{Received \today; accepted XXXX}

% \abstract{}{}{}{}{} 
% 5 {} token are mandatory
 
  \abstract
   {Observations of slow magnetoacoustic waves in solar coronal loops suggest that, in hot coronal plasma, heat conduction may be suppressed in comparison with the classical thermal transport model. }
   {We link this suppression with the effect of the non-local thermal transport that appears when the plasma temperature perturbation gradient becomes comparable to the electron mean free path. Moreover, we consider a finite time of thermalisation between electrons and ions, so that separate electron and ion temperatures can occur in the loop.}
   {We numerically compare the influence of the local  and non-local 
thermal transport models on standing slow waves in one- and two-temperature coronal loops. To quantify our comparison, we 
use the period and damping time of the waves as commonly observed parameters.}
   {Our study reveals that non-local thermal 
transport can result in either shorter or longer slow-wave damping times in comparison with the local conduction model due to the 
suppression of the isothermal regime. The difference in damping times can reach 80\%. For hot coronal loops, we found that the 
finite equilibration between electron and ion temperatures results in up to 50\% longer damping time compared to the one-temperature 
case. These results indicate that non-local transport will influence the dynamics of compressive waves across a broad range of 
coronal plasma parameters with Knudsen numbers (the ratio of mean-free-path to temperature scale length) larger than 1\%.}
   {In the solar corona, the non-local thermal transport shows a significant influence on the dynamics of standing slow waves in a broad range of plasma parameters, while two-temperature effects come into play for hot and less dense loops.}


   \titlerunning{Non-local transport impact on compressive waves}
   \authorrunning{S.A. Belov et al.}
   
   \maketitle
%
%-------------------------------------------------------------------
\newcommand{\Freyja}{{\sf Freyja}}
\let\oldpageref\pageref
\renewcommand{\pageref}{\oldpageref*}

\section{Introduction}
\label{sec:intro}
Standing slow waves in the solar corona were first observed as periodic variations of the Doppler shift in hot lines \citep{Wang2002, 
Wang2003, Mariska2005, Mariska2006}. These oscillations are called SUMER oscillations after the Solar Ultraviolet Measurement of Emitted 
Radiation (SUMER) spectrometer onboard Solar and Heliospheric Observatory. Later, standing slow waves associated with SUMER oscillations 
in flaring coronal loops were detected by imaging observations \citep{Wang2015, Nistic2017}. The intrinsic feature of SUMER oscillations 
is a rapid decay with characteristic times comparable to their periods. These damping times can be used as seismological information to 
determine the transport coefficients in coronal plasma \citep{Wang2018, Wang2019} and constrain the coronal heating function 
\citep{2019ApJ...884..131R, 2020A&A...644A..33K}.

At the same time, there is evidence of standing slow waves existing in cooler coronal loops obtained by spectroscopic and imaging 
instruments. Using EUV imaging-spectrometer Hinode/EIS, \cite{Erdlyi2008} analysed oscillations of intensity and Doppler shift along 
coronal loop-like structures with 1-2\,MK temperatures. Due to the quarter-period phase shift between the intensity and the Doppler shift, 
these oscillation were explained as standing slow waves. Using the same instrument, for Fe XII, Fe XIII, Fe XIV and Fe XV coronal lines, 
\cite{Mariska2008} observed the Doppler shift oscillations with the amplitude of about 2 km/s and period of around 35 minutes consistent 
with the explanation in terms of standing slow waves. For these observations, the damping time increased with increasing temperature: for the Fe XII line (the lowest formation temperature), the damping time was 42 minutes, while, for the Fe XV line (the highest formation temperature), no damping was observed.
The latter, the lack of slow-wave damping, was attributed to the effect of thermal misbalance in \citet{Kolotkov2019}.
For coronal fan loops, \cite{Pant2017} reported imaging observations 
of standing slow wave with period and damping time of about 25 and 45 minutes, respectively.

Statistical analysis made by \cite{Nakariakov2019} shows that there is a gap in observations of standing slow waves between hot coronal 
loops with temperatures 6--14\,MK, where these waves are predominantly observed, and cold/warm coronal loops with temperatures 0.6--2\,MK. For hot coronal loops, one may expect the field-aligned thermal conduction to be the main actor for the slow wave dynamics. However, the seismologic measurement 
of the polytropic index using a standing slow wave in a 10\,MK loop showed a value near 1.64 \citep{Wang2015}. Since this value is close 
to the adiabatic
index 5/3 for an ideal monatomic gas, \cite{Wang2015} suggested that the thermal conduction should be at least 3 times suppressed in the hot 
ﬂare loops. For 1.7\,MK plasma, \cite{VanDoorsselaere2011} estimated the polytropic index to be 1.10 using slow waves propagating upward 
from active region footpoints. The measured value is close to the isothermal adiabatic index of 1, which suggests that thermal conduction is 
important in these conditions. Moreover, unlike \citep{Wang2015}, the thermal conduction value inferred from the measured phase shift was 
comparable to the classical value \citep{Spitzer1953}. On the other hand, investigating propagating slow waves in fan loops,  
\cite{Prasad2018} found that the polytropic index varied from 1.04 to 1.58 being higher for hotter loops while the temperature range 
considered was approximately 0.9--1.1\,MK. It should be mentioned that all these inferences on the role of conduction
were made in the polytropic assumption. It was 
shown by \citet{Kolotkov2022} that this assumption can be used in the corona only in a weakly conductive regime when the effective polytropic index 
weakly deviates from its adiabatic value of 5/3.
%In the opposite case, its estimation should be done as a ratio between observed and expected (in the adiabatic case) periods.

When fitting polytropic models with conduction to the damping of slow waves, the higher values of the polytropic index inferred for 
hotter loops are in contradiction to the classical thermal conduction theory \citep{Spitzer1953} which would predict a decrease in the 
inferred polytropic index with temperature \citep[see e.g. Fig. 5 from][]{Kolotkov2022}. The heat fluxes predicted by this local thermal
transport model depend on the local temperature gradient. It assumes that the thermal electron mean-free path is much less than the 
characteristic length scales of the temperature gradient. However, as the temperature increases, the mean-free path can become comparable
with the characteristic length-scales and hot electrons from the tail of the electron distribution can travel for longer distances 
between effective collisional scattering. This makes the local thermal transport model invalid and demands using a non-local model based
on solving the Vlasov-Fokker-Planck equation \citep{Arber2023}. The main effects described by the non-local models of thermal transport are the broadening of heat flux effective spatial profile (also referred to as preheating, caused by hotter electrons carrying energy for longer distances) and an
overall suppression of heat fluxes. In other words, non-local thermal transport models allow for mitigating the infinite increase of the heat flux with temperature in the classical thermal conduction theory by \citet{Spitzer1953}.
In the case of coronal loops, this suppression may result in an increase in coronal cooling times 
\citep{West2004ESASP.575..573W, West2008}. This suggests that non-local thermal transport may explain the suppression of thermal 
conduction observed by using slow waves \citep{Wang2015, Prasad2018} and in solar flare observations \citep{Battaglia2009, Fleishman2023}. However, the influence of the non-local thermal effects on slow 
waves has yet to be studied.

In this work, we compare the influence of the local and non-local thermal transport models on standing compressive waves for a broad 
range of coronal plasma conditions. Also, we examine how finite equilibration between ion and electron temperatures affect the wave 
dynamics. The paper is organised as follows. In Section~\ref{sec:num_model}, we describe the numerical code and the thermal transport 
models used in this study. In Section~\ref{sec:simulations}, we conduct a parametric study of the thermal transport influence on standing 
slow waves in one- and two-temperature coronal loops. Finally, we draw our conclusions in Section~\ref{sec:conclusions}.

\section{Numerical model}
\label{sec:num_model}

\begin{figure*}
\includegraphics[width=\textwidth]{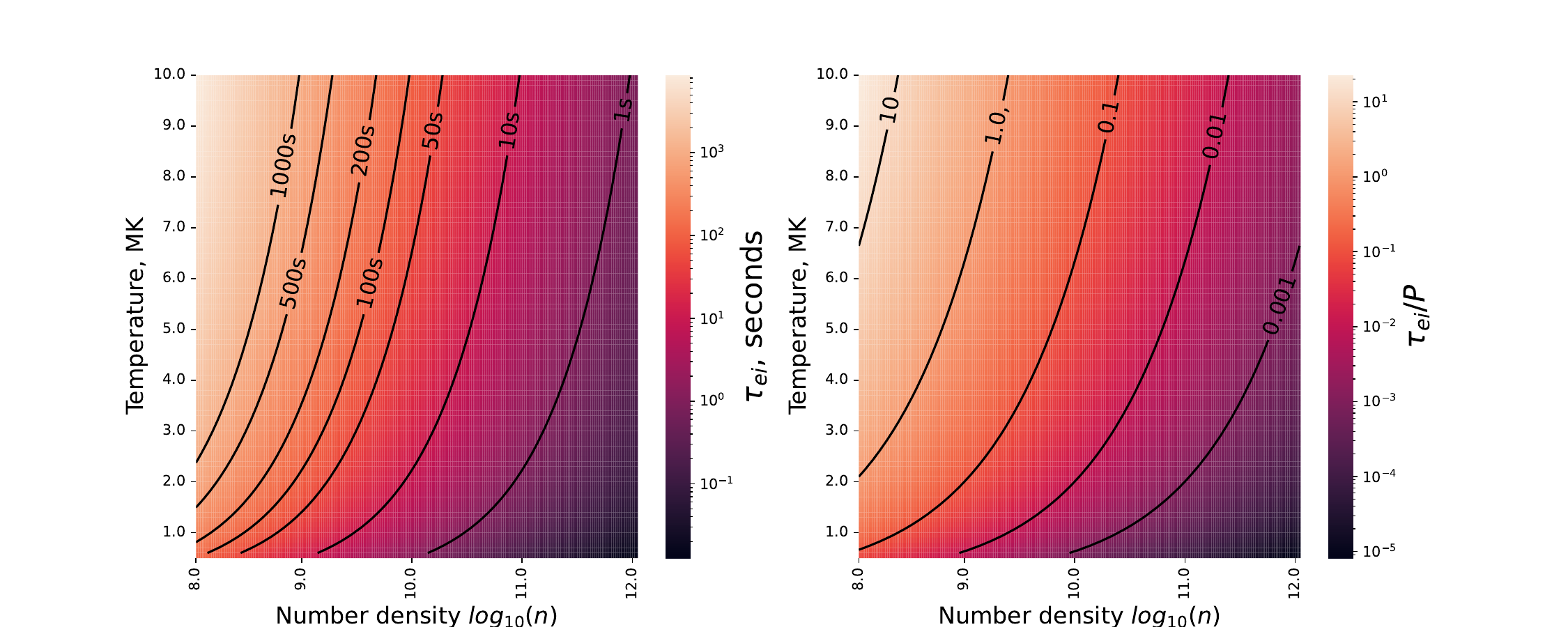}
    \caption{Left panel: equilibration time $\tau_{ei}$ contours as a function of the plasma number density and temperature. Right panel: 
    the relation between the equilibration time $\tau_{ei}$ and a period $P$ of a fundamental standing mode in a 100\,Mm loop in adiabatic plasma 
    as a function of the plasma number density and temperature.}
    \label{fig:eq_time}
\end{figure*}

\begin{figure*}
\includegraphics[width=\textwidth]{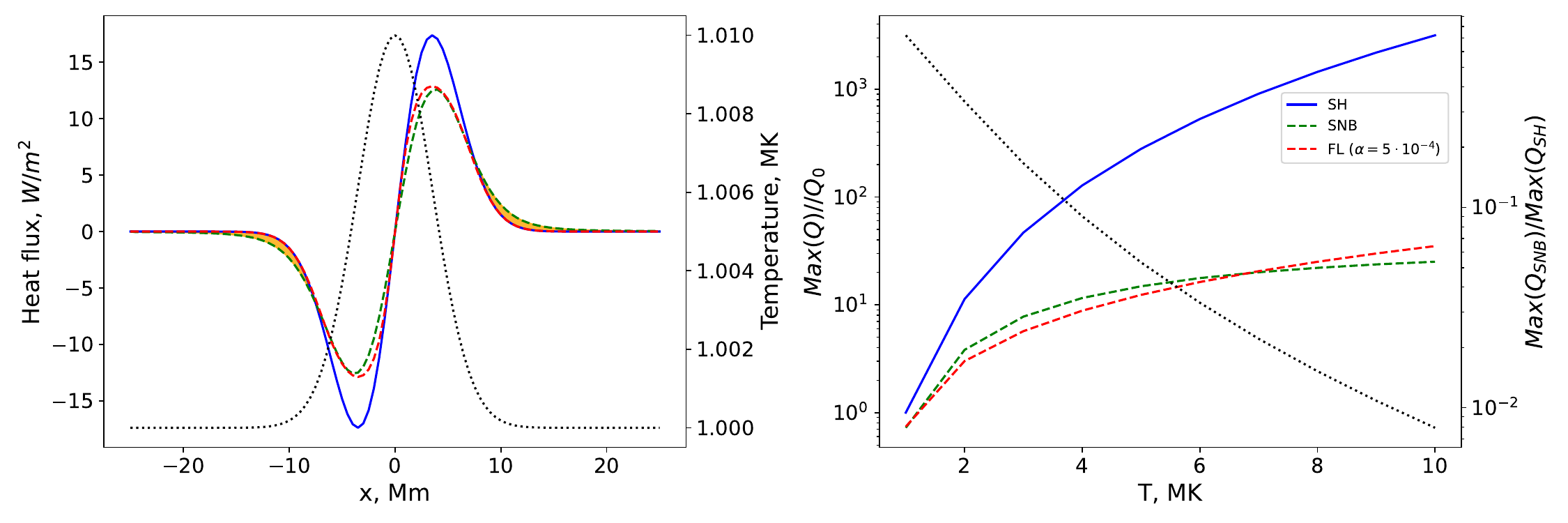}
    \caption{Left panel: Initial Gaussian perturbation of temperature (black dotted curve) and heat fluxes corresponding to this 
    perturbation for the case of SH (blue solid curve), SNB (green dashed curve) and FL (red dashed curve) thermal transport models. The orange shading denotes the regions where pre-heating takes place (the effective width of the heat flux predicted by SNB is greater than that by SH and than the width of the initial temperature perturbation). Right panel: the peak SH (blue solid curve), SNB (green dashed curve) and FL (red dashed curve) heat fluxes measured in units of maximum SH heat flux at 1MK ($Q_0$); the black dotted curve denotes the relation of maximum SNB heat flux over  the corresponding maximum SH heat flux.}
    \label{fig:fluxes}
\end{figure*}

In this study, we use the \Freyja~code -- a two-temperature hydrodynamics code designed originally for simulating laser-plasma 
experiments \citep[e.g.][]{paddock2023energy}, which solves the equations for mass and momentum conservation coupled to non-adiabatic
equations for the specific internal energies of the ion and electron species of a plasma:

\begin{equation}
    \label{eq:cont}
    \frac{D \rho}{D t} + \rho \nabla \cdot \vec{v} = 0,
\end{equation}

\begin{equation}
    \label{eq:motion}
    \rho \frac{D \vec{v}}{D t} + \nabla (P_T + Q_{\mathrm{Visc}}) = 0,
\end{equation}

\begin{equation}
    \label{eq:energy_ion}
    \rho \frac{D \epsilon_i}{D t} + (P_i + Q_{\mathrm{Visc}}) \nabla \cdot \vec{v} = S_{\mathrm{equil}},
\end{equation}

\begin{equation}
    \label{eq:energy_ele}
    \rho \frac{D \epsilon_e}{D t} + P_e \nabla \cdot \vec{v} = S_{\mathrm{equil}} + S_{\mathrm{Cond.}}.
\end{equation}

Here, we use the hydrodynamical model without taking magnetic field perturbations into account because in the low-$\beta$ plasma slow waves result in the perturbation of the magnetic field much smaller than the density and temperature perturbations \citep[e.g.][]{Ofman2022}. This 1D approach for modeling slow waves in the corona is also known as the infinite field approximation \citep[see][Sec.~2.3 and references therein]{2021SSRv..217...34W}.

The core hydrodynamics equations are solved in the Lagrangian frame, using a compatible energy update \citep{caramana1998construction}.
$Q_{\mathrm{Visc}}$ represents a shock viscosity \citep{caramana1998formulations}. The viscous heating only impacts the ion equation, and
$P_T = P_i + P_e$ is the total hydrodynamical pressure. The specific internal energies $\epsilon_{i,e}$ are the internal energies of the 
ions and electrons, divided by the total (electron and ion) mass density. The plasma is assumed to be fully ionised hydrogen, with an ideal gas equation of state,
\begin{equation}
    \label{eq:state}
    P_{e,i} = (\gamma - 1) \rho  \epsilon_{e,i},
\end{equation}
where $\gamma = 5/3$.

The term $S_{\mathrm{equil}}$ represents the change in temperature due to collisional temperature equilibration of 
the ion and electron species. The rate of change of the electron temperature is given by:
\begin{equation}
    \label{eq:temp_equil}
    \frac{dT_e}{dt}=\frac{1}{\tau_{ei}}\left(T_i-T_e\right),
\end{equation}
where the equilibration time $\tau_{ei}$ is given by the inverse of the collision frequency:

\begin{align}
  \label{eq:equil_time}
  \nu^{ei} &= \frac{q_e^4}{3.0 k_B^{3/2} \sqrt{2.0} \pi^{1.5}  \epsilon_0^2}\nonumber \frac{\left(m_e m_i\right)^{1/2} Z_e^2 Z_i^2 n_i \lambda_{e i}}
     {\left(m_e T_i + m_i T_e\right)^{3/2}}\\
  &\approx 6.94 \times 10^{-21} \frac{\left(m_e m_i\right)^{1/2} Z_e^2 Z_i^2 n_i \lambda_{e i}}
     {\left(m_e T_i + m_i T_e\right)^{3/2}},
\end{align}
where $m_{e,i}$, $n_{e,i}$, $T_{e,i}$, $Z_{e,i}$ are masses, number densities, temperatures, and charge numbers of electrons and ions, respectively; $k_B$ is the Boltzmann constant, and $\lambda_{ei}$ is the electron-ion Coulomb logarithm. To justify the importance of treating electron and ion temperatures separately in our study, we plotted the equilibration time $\tau_{ei}$ in the left panel of 
Fig.~\ref{fig:eq_time} for a wide set of coronal densities and temperatures. This plot shows that this time may vary from a fraction of a
second to several thousand seconds. The right panel of Fig.~\ref{fig:eq_time}, compares this time with the period $P$ of the fundamental 
standing slow mode in an adiabatic plasma calculated as $P=2L_0/c_s$, where $L_0=100$\,Mm is the loop length, $c_s$ is the adiabatic sound 
speed. It can be  seen that there is a region where $\tau_{ei}$ is comparable with wave periods and, therefore, this effect should be 
taken into account in the study.

In Eq.~(\ref{eq:energy_ele}), $S_{\mathrm{Cond.}}$ represents the heat flux:

\begin{equation}
    \frac{\partial \epsilon_e}{\partial t} = - \nabla \cdot \vec{q}.
\end{equation}

In this work, ion thermal conduction is neglected. The electron heat flux, $\vec{q}$ can be calculated either as a heat flux obtained 
for short mean-free-paths by using Spitzer-H\"{a}rm (SH) approximation \citep{Spitzer1953} $Q_{\mathrm{\mathrm{SH}}}$, the flux-limited  (FL) local heat flux $Q_{\mathrm{\mathrm{FL}}}$ or the non-local heat flux  obtained with the Schurtz-Nicolaï-Busquet (SNB) model \citep{Schurtz2000, Brodrick2017} $Q_{\mathrm{\mathrm{SNB}}}$.

The SH heat flux is calculated as follows:
\begin{align}
    \label{eq:SH_flux}
    &Q_{\mathrm{\mathrm{SH}}} = -\kappa_0 T_e^{5/2} \nabla T_e,\\ 
    &\kappa_0 \approx 5.76\times10^{-11}\frac{\Gamma\left(Z_i\right)}{Z_i \lambda_{ei}},\nonumber \quad
    \Gamma\left(Z_i\right) = 13.6\frac{\left(Z_i+0.24\right)}{\left(Z_i+4.24\right)}.\nonumber
\end{align}
For fully-ionised hydrogen plasma $Z_i=1$ and 
$\lambda_{ei}=18.5$ is frequently assumed for the coronal conditions, giving $\kappa_0\approx10^{-11}$. While this SH approximation
is common in solar coronal studies it is only a valid approximation for Knudsen numbers $K_n=\lambda_{ei}^T/\lambda_T<0.01$, where $\lambda_{ei}^T$ is the electron thermal mean-free path, and $\lambda_T$ is the temperature 
gradient length scale.

In regions of steep temperature gradients, the predicted SH flux can become unphysical. For example, the total heat flux is unlikely to
be larger than the total local thermal energy density advected at the thermal speed, the so-called
free-streaming limit heat flux $q_{fs} = v_{th} n_e k_B T_e$. As a result the heat
flux is often limited to make sure it is less than this limit.
The FL local heat flux is calculated by applying a Larsen limiter \citep{OLSON2000619} to the Spitzer-H\"{a}rm flux, 
\begin{align}
    \label{eq:FL_flux}
    &Q_{\mathrm{FL}} = -\kappa'\nabla T_e,\\ 
    &\kappa' = \frac{1}{\sqrt{\dfrac{1}{\left(\kappa_0 T_e^{5.2}\right)^2} + \left(\dfrac{1}{\alpha q_{fs}}\left|\nabla 
    T_e\right|\right)^2}}, \nonumber \\
    &q_{fs} = v_{th} n_e k_B T_e, \quad
    v_{th} = \sqrt{\frac{2 k_B T_e} {m_e}} \nonumber,
\end{align}
where $\alpha$ is the maximum fraction of the free-streaming flux permitted. Typically this is between 0.01 and 0.2 and is usually determined by the best fit to observations. This limits the predictive power of this approach.

The heat flux in the SH model is primarily carried by electrons with speeds $\simeq 3 v_{th}$. Since the mean-free-path scales with 
speed to the fourth power such electrons have mean-free-paths $\simeq 100$ times larger than the thermal mean-free-path. Hence SH is
expected to be inaccurate once $K_n>0.01$. A full treatment of heat conduction in this regime requires a Vlasov-Focker-Planck (VFP) 
solution.
This is too complex to be included in a fluid code and instead a simplified model, the SNB model, is often used.
The SNB model is an approximate solution of the full VFP equation divided into energy bins \citep[for the 
derivation outline see ][]{Schurtz2000, Brodrick2017}. These energy bins are introduced to approximate the distribution function. The 
simplified VFP equation is solved for each energy bin, and the solutions are combined to calculate a correction to the local SH heat flux $Q_{\mathrm{\mathrm{SH}}}$.
The non-local heat flux obtained with the SNB model \citep{Schurtz2000, Brodrick2017}, is
\begin{equation}
    \label{eq:SNB_flux}
    Q_{\mathrm{\mathrm{SNB}}} = Q_{\mathrm{\mathrm{SH}}} - \sum_{g}{\frac{\lambda_g}{3}\nabla H_g}, 
\end{equation}
where the function $H_g$ is calculated by solving ODE (15) from \citet{Arber2023} for each energy group $E_g$ with $\lambda_g^2=\lambda_{ee}\lambda_{ei}$, where $\lambda_{ee}$ and 
$\lambda_{ei}$ are electron-electron and electron-ion mean-free paths \citep[for the details see][]{Arber2023}. 

To illustrate the differences between the aforementioned thermal transport models, we calculated the heat fluxes for the initial Gaussian perturbation of plasma temperature in a form $T=T_0 + 0.01T_0 \exp\left(-x^2/\sigma^2\right)$, where $T_0=1$\,MK, 
$\sigma=5$\,Mm and plasma number density $n_0=10^{9}$\,cm$^{-3}$, by using Eqs.~(\ref{eq:SH_flux}), (\ref{eq:FL_flux}), and (\ref{eq:SNB_flux}). The SNB heat flux was calculated by adopting the python script from \citet{Arber2023}\footnote{The Python script and Jupyter notebook can be accessed on GitHub: https:\url{//github.com/Warwick-Solar/SNB_flux}}.  The calculated heat fluxes are shown in the left panel of Fig.~\ref{fig:fluxes}. It can be seen that SNB results in a suppression of the heat
flux and its spreading for a wider region compared to local SH and FL models. This pre-heat is due to the high energy tail of the 
electron distribution. Thus SH over-estimates the heat flux for $K_n>0.01$. This reduction in heat flux can be approximated using the
FL model but only once the answer is known so that $\alpha$ can be specified. The SNB model has no free parameters and predicts
both flux suppression and pre-heat.

With an increase in temperature, the maximum heat flux value increases for all models considered (see blue, green, and red curves in the 
right panel of Fig.~\ref{fig:fluxes}). However, this growth is significantly slower for the FL and SNB models. Moreover, the results for the FL model are subject to the choice of the parameter $\alpha$. For the SNB case, the effective reduction of heat flux in 
comparison to the SH model is connected with the increase in the electron mean-free-path with temperature. The black curve in the right 
panel of Fig.~\ref{fig:fluxes} demonstrates that $Q_{\mathrm{\mathrm{SNB}}}$ decreases with increasing temperature from $Q_{\mathrm{\mathrm{SNB}}}\approx 0.7 Q_{\mathrm{\mathrm{SH}}}$ at 
$T_0=1$\,MK to $Q_{\mathrm{\mathrm{SNB}}}\approx 0.008 Q_{\mathrm{\mathrm{SH}}}$ at $T_0=10$\,MK.

\section{Numerical simulations}
\label{sec:simulations}

% \subsection{Gaussian temperature perturbation}

\subsection{Slow wave mode set up}
\label{sec:setup} 

\begin{figure*}
\includegraphics[width=\textwidth]
{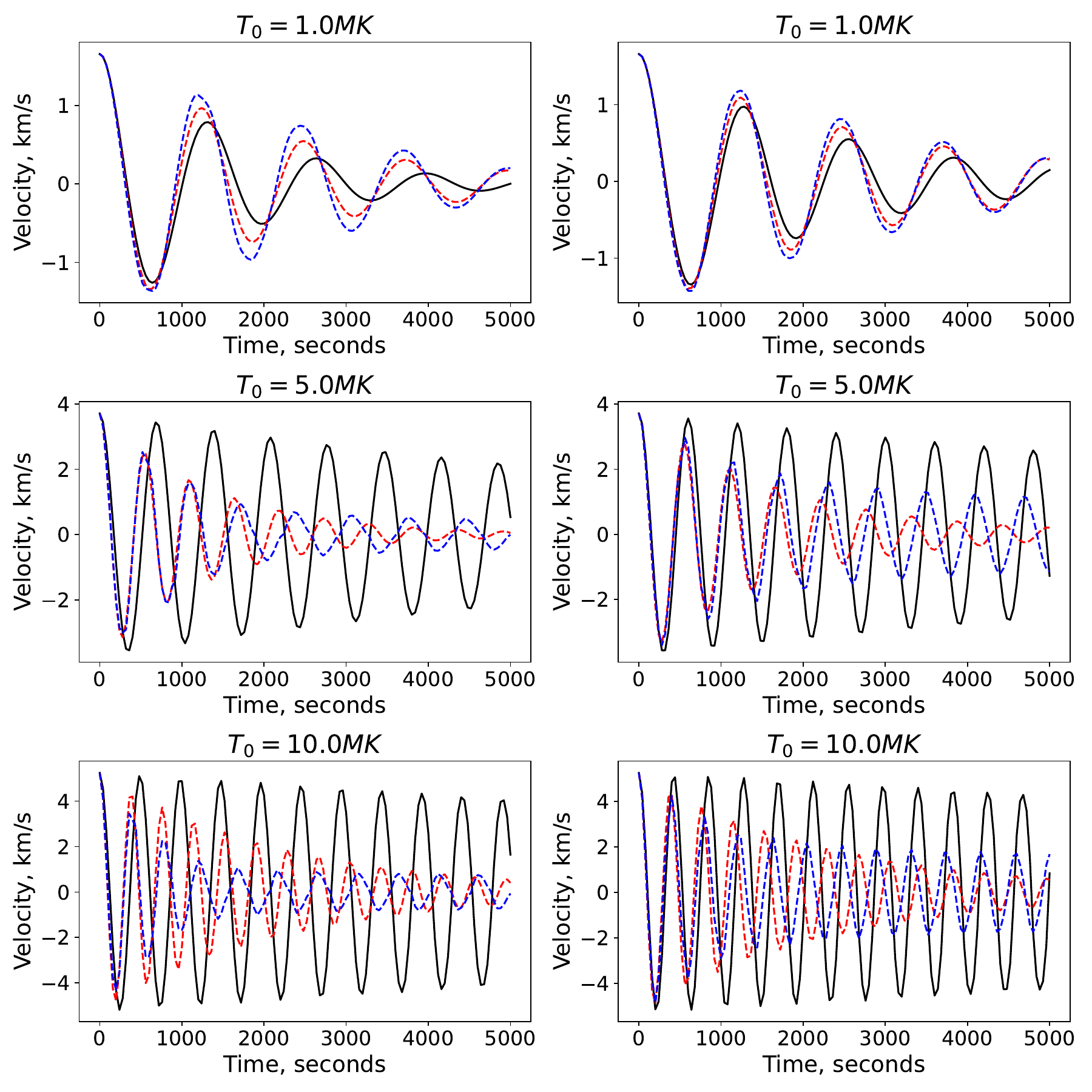}
    \caption{Examples of velocity perturbations at the apex of a loop with $L=100$\,Mm and $n_0=10^8$\,cm$^{-3}$  for SH (black solid curve),
    SNB (red dashed curve), and FL with $\alpha=0.0001$ (blue dashed curve) in one-temperature (left column) and two-temperature (right 
    column) regimes.}
    \label{fig:apex_vel}
\end{figure*}

To simulate standing slow waves, we use 1000 uniformly spaced grid points and reflecting boundary conditions assuring a zero thermal flux through the 
boundaries. The shock viscosity is switched off since all the simulations are linear and do not involve any shocks. To initialise the fundamental standing mode, we set up the initial velocity perturbation as:
\begin{equation}
\label{eq:vel_pertrub}
    v = 0.01 c_{s} \sin\left(\frac{\pi x}{L_0}\right),
\end{equation}
where $c_s$ is a sound speed, and $L_0$ is a length of the loop, which is equal to the size of the domain. All runs are performed for a set of loop parameters listed in 
Table~\ref{tab:parameters} using the following thermal transport models: SH, FL, and SNB. For the SNB model, we use 50 energy bins 
uniformly spread between 0 and $20k_B T_0$ to assure both adequate energy resolution and computational speed.

Fig.~\ref{fig:apex_vel} shows examples of the velocity evolution at the loop apex for the considered models of thermal transport in one- 
(left panel) and two- (right panel) temperature regimes. It follows from the plot that the two-temperature regime results in slower 
damping of standing waves. Also, it can be seen that the difference between non-local and local thermal transport is not negligible. At 
the same time, the SNB and FL models almost coincide at $T_0=1$\,MK for the chosen value of the flux-limiting parameter $\alpha=0.0001$ and
diverge with the increase in temperature. It demonstrates that the choice of the the flux-limiting parameter $\alpha=0.0001$ is problem-dependent and it is impossible to use the same value of $\alpha$ for all domains of the plasma considered. Moreover, at $T_0=10$\,MK, the FL
model shows apparent non-exponential damping (the oscillation amplitude approaches a stationary level), while the damping is exponential in the SH and SNB cases. This  non-exponential 
damping is attributed to the form of flux-limiting and, therefore, is non-physical.  For the reasons mentioned above, we focus on the comparison between the SH and SNB models.

\begin{table*}
	\centering
	\caption{Set of the loop parameters used in the study.}
	\label{tab:parameters}
	\begin{tabular}{cc}
		\hline
		Parameter & Value\\
		\hline
		Loop length $L_0$, Mm  & $100$ \\
		Loop temperature $T_0$, MK & $0.5-10.0$ with $0.5$ step  \\
		Decimal log of loop number density (expressed in cm$^{-3}$) $\log_{10}\left(n_0\right)$ & $8.0-11.0$ with $0.5$ step \\
		\hline
  
	\end{tabular}
\end{table*}

\subsection{One-temperature (MHD) regime}
\label{sec:one_t_reg} 
\begin{figure}
\includegraphics[width=\linewidth]{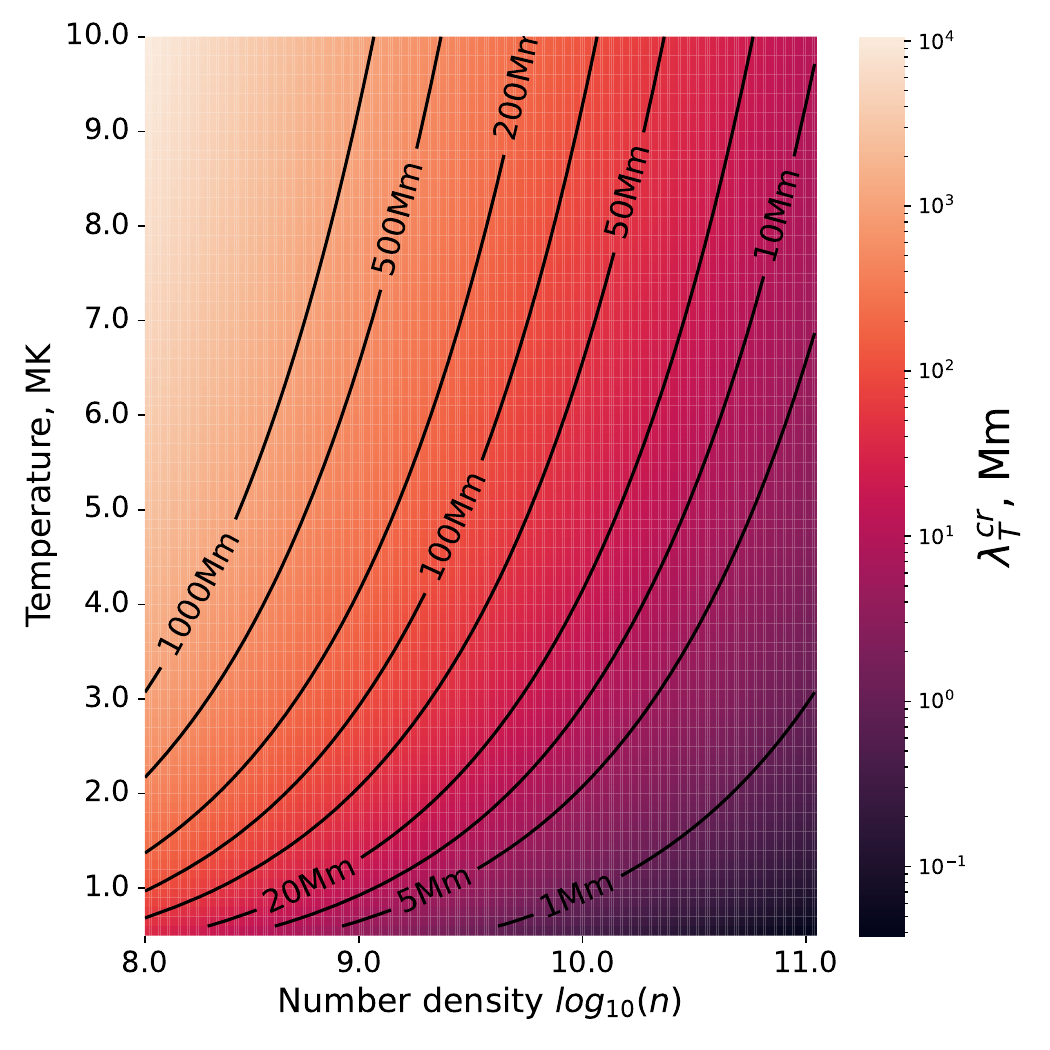}
    \caption{Critical temperature length scale $\lambda_{T}^{cr}$ as a function of the plasma density and temperature, below which the non-local thermal transport effects are required.}
    \label{fig:knudsen}
\end{figure}

To initially separate the effects of non-local thermal transport from the effect of the finite equilibration time $\tau_{ei}$, we multiply the value of $\tau_{ei}$, calculated using Eq.~(\ref{eq:equil_time}), by $10^{-20}$, effectively making temperature equilibration instantaneous. This guarantees that plasma is in the MHD (one-temperature) regime. Next,  to separate the local thermal transport regime from the non-local one, it is worth using Knudsen number $K_n=\lambda_{ei}^T/\lambda_T$. Usually, the temperature length scale is determined as $\lambda_T = T\left(dT/dx\right)^{-1}$, however, in this work, we use  $\lambda_T = L_0$ 
since we study small perturbations of uniform background and may underestimate $\lambda_T$. It was shown that for $K_n > 0.01$ the local thermal transport 
approximation fails and non-local effects take place \citep{Arber2023}.  Fig.~\ref{fig:knudsen} demonstrates the critical temperature 
gradient scale length $\lambda_T^{cr}=0.01\lambda_{ei}^T$, for $\lambda_T<\lambda_T^{cr}$ the local thermal transport may be violated. 
For example, for hot coronal plasma $\lambda_T^{cr}\approx200$\,Mm which is comparable with the loop lengths observed. 

 To check the difference between the local and non-local thermal transport models, we use the velocity profiles at the middle of numerical domain (as in Fig.~\ref{fig:apex_vel}) and fit them by a decaying harmonic function
 $v\left(x=0.5L_0, t\right) \propto e^{- t/\tau}\sin\left({2\pi t}/{P} + \phi_0\right)$ with initial phase $\phi_0$ to retrieve the oscillation period $P$ and damping time $\tau$ as potential observables.
%with a polynomial trend:
%\begin{equation}
%    v\left(x=0.5L_0, t\right) = a_0 e^{- t/\tau}\sin\left(\frac{2\pi t}{P} + \phi_0\right) + c_2  t^2 + c_1  t + c_0. 
%	\label{eq:fit_function}
%\end{equation}
A similar approach was used by \cite{Wang2015} to retrieve the parameters of slow-mode waves in a flaring loop.
%We chose the slow wave period $P$ and damping time $\tau$ because these parameters are observable parameters of standing slow waves in coronal loops.
Thus, using these parameters, we can estimate how the different thermal transport models affect slow wave dynamics for
the different coronal plasma conditions. 

\begin{figure*}
\includegraphics[width=\textwidth]{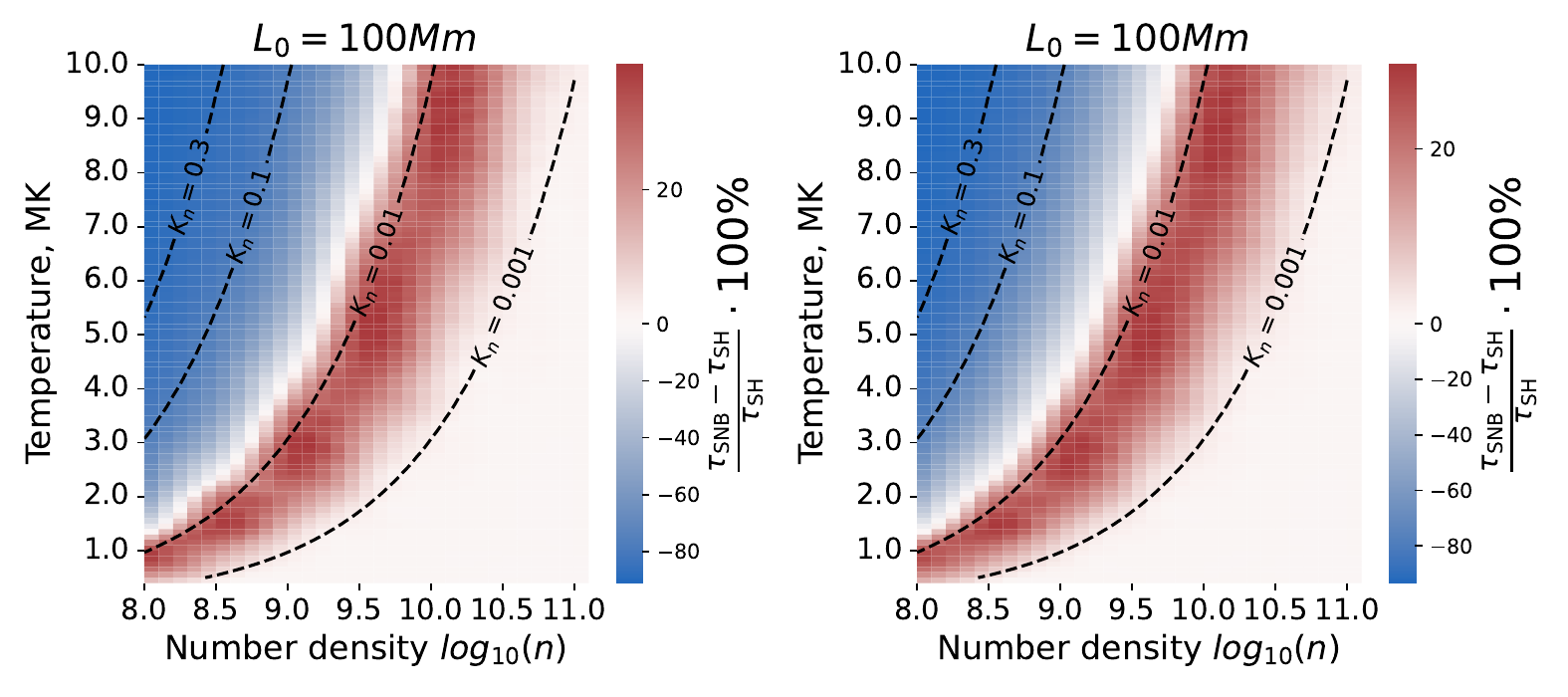}
    \caption{Relative difference between SNB and SH damping times of slow waves for the one- (left panel) and two-temperature (right 
    panel) plasma regimes. The plot was obtained by using cubic interpolation of the results from $0.5\times0.5$ grid cell to 
    $0.1\times0.1$ grid cell.}
    \label{fig:comparison_tau}
\end{figure*}

\begin{figure*}
\includegraphics[width=\textwidth]{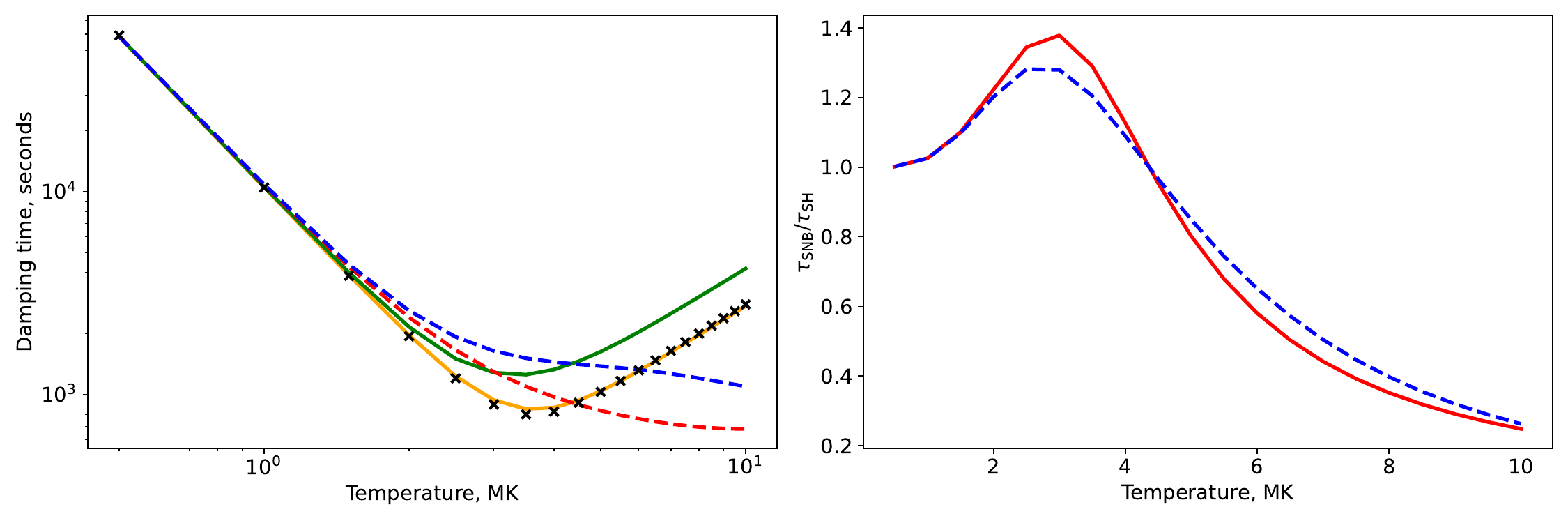}
    \caption{Left panel: Slow waves damping times for SH in one- (orange solid curve) and two-temperature (green solid curve) regimes, 
    and SNB in one- (red dashed curve) and two-temperature (blue dashed curve) in plasma with $n_0=10^{9}$\,cm$^{-3}$. The black crosses 
    represent the damping time calculated by solving a dispersion relation \citep[see e.g. Eq.~(19) in ][]{2003A&A...408..755D} for one temperature plasma with SH thermal transport model. 
    Right panel: ratio of SNB and SH damping times for one- (red solid curve) and two- (blue dashed curve) temperature plasma with
    $n_0=10^{9}$\,cm$^{-3}$.}
    \label{fig:taus}
\end{figure*}

\begin{figure*}
\includegraphics[width=\textwidth]{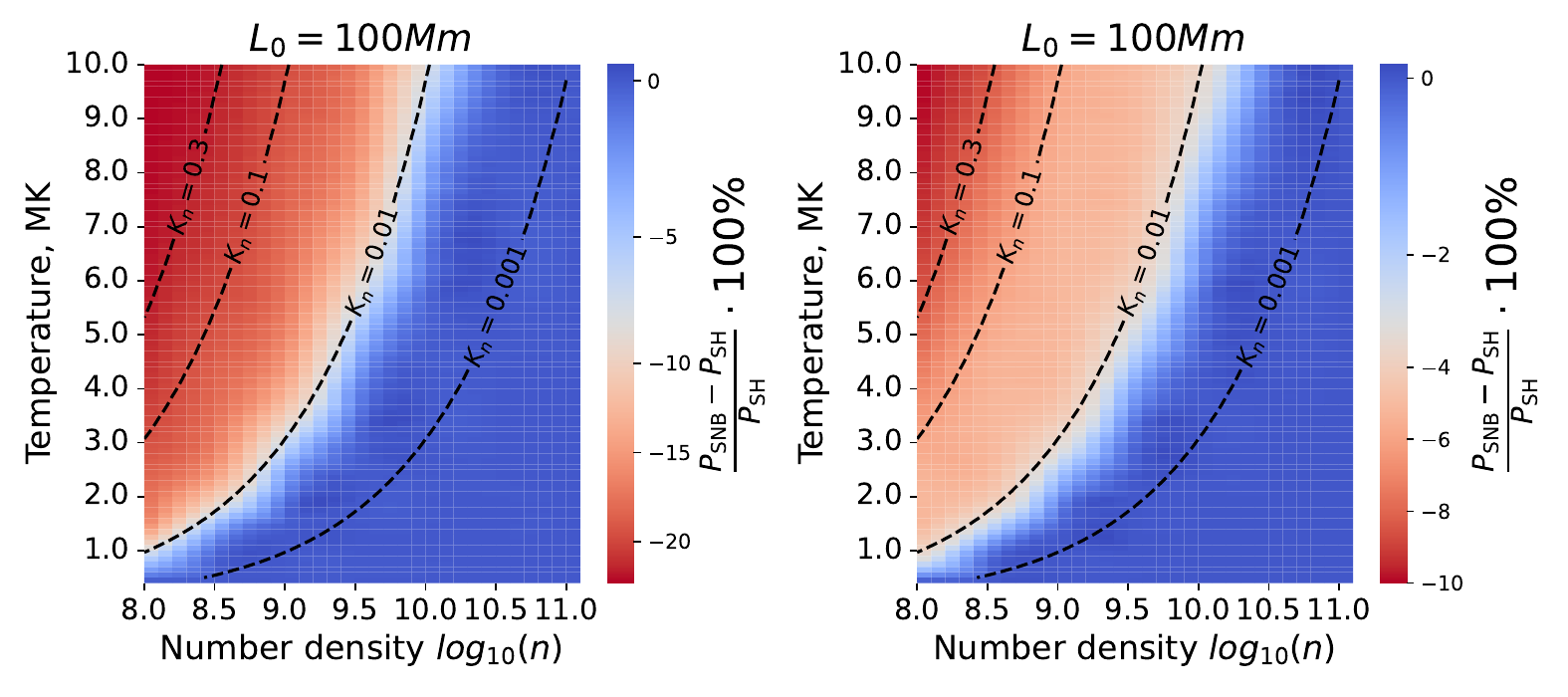}
    \caption{Relative difference between SNB and SH periods of slow waves for the one- (left panel) and two-temperature (right panel) 
    plasma regimes. The plot was obtained by using cubic interpolation of the results from $0.5\times0.5$ grid cell to 
    $0.1\times0.1$ grid cell.}
    \label{fig:comparison_p}
\end{figure*}

The left panel of Fig.~\ref{fig:comparison_tau} shows the relative difference between the damping time of slow waves with non-local 
thermal transport $\tau_{\mathrm{SNB}}$ and the damping time for the local thermal transport model $\tau_{\mathrm{SH}}$ together with 
$K_n$-contours for the one-temperature regime for coronal loops with $L_0=100$\,Mm \citep[cf. Fig. 6 for solar flares in][]{Battaglia2009}. As one can see from Fig.~\ref{fig:comparison_tau}, the damping of slow waves with the SNB model can be both more and less effective than with the SH model, and the absolute value of the 
difference between damping 
times can reach 80\%. The border between the regimes of faster/slower damping is found to be near $K_n=0.01$. Since non-local thermal 
transport results in thermal flux suppression, faster damping of slow waves seems to be in contradiction in this case. 
However, this
can be explained by the left panel of Fig.~\ref{fig:taus} showing the dependencies of SNB and SH damping times on temperature for plasma
number density $n_0=10^{9}$\,cm$^{-3}$.
To validate the results of our numerical simulations, Fig.~\ref{fig:taus} also shows the dependence of the SH damping time on temperature, obtained from the dispersion relation of slow-mode wave in one-temperature plasma with thermal conduction \citep[see e.g. Eq.~(19) in ][]{2003A&A...408..755D}.
Thus, for the SH case, the damping time curve reaches a minimum near 3\,MK. Up to this value, increasing temperature results in an increase in thermal conductivity and consequent reduction in the damping time due to conduction.
However, for very high temperatures conduction is so rapid that the oscillations approach the isothermal limit. In this regime, conduction
can have no effect on the damping of the slow mode and thus the damping time increases. 
This explains the increase in damping time after the minimum (see also the left panel of Fig. \ref{fig:apex_vel}). Thus, we can 
conclude that for the SH model, the waves gradually approach an isothermal regime after 3\,MK. In the SNB case, in contrast, the heat-flux suppression due
to non-local thermal transport results in the damping time minimum being shifted to higher temperatures ($\gtrsim 10$\,MK according to Fig.~\ref{fig:taus}). Therefore, SNB non-local conduction  
causes a delayed onset of the isothermal regime compared to SH.

The right panel of Fig.~\ref{fig:taus} shows how the ratio between SNB and SH damping times changes with temperature for plasma number density 
$n_0=10^{9}$\,cm$^{-3}$. It can be seen that the deviation between SNB and SH damping times is non-monotonic with the increase in temperature corresponding to 
$K_n$ increase. Initially, the SNB damping time grows in comparison with the SH case and drops down as soon as waves approach the isothermal regime in the SH 
case.

The left panel of Fig.~\ref{fig:comparison_p} represents the relative difference of measured periods of slow standing waves for the SNB ($P_{\mathrm{SNB}}$) and the SH ($P_{\mathrm{SH}}$) models for one-temperature regime as it was shown for the damping time $\tau$ in Fig.~\ref{fig:comparison_tau}. The region 
where the difference is noticeable starts from the $K_n=0.01$ contour. The difference is negative inside this domain. It means that the period of standing waves 
is shorter for the case of non-local thermal transport and, therefore, their phase speed is higher. For the local thermal transport case, the phase speed of 
slow waves decreases from the usual sound speed to the isothermal sound speed. As the non-local thermal transport prevents slow waves from switching to the 
isothermal regime, the phase speed of slow waves decreases more slowly than in the case of local thermal transport and, thus, yields a negative difference in 
periods. In this regard, non-local transport allows the periods of slow waves to remain closer to their values in the adiabatic plasma.

Thus, our parametric study for the one-temperature (MHD) regime showed that the suppression of heat fluxes by non-local transport leads to the later 
appearance of the isothermal regime. From the observational point of view, it results in higher/smaller damping times and shorter, close to adiabatic values, 
periods of standing slow waves. Moreover, the difference in damping time can reach 80\%.

\subsection{Two-temperature regime}
\label{sec:two_t_reg} 
As shown in Fig.~\ref{fig:eq_time}, the value of the equilibration time $\tau_{ei}$ can be comparable with the observed periods of slow waves in the solar corona. In 
this case, the temperatures of electrons and ions are not strongly coupled and plasma exists in the two-temperature regime. The right panels of 
Fig.~\ref{fig:comparison_tau} and Fig.~\ref{fig:comparison_p} reflect the same as their left panels but for the two-temperature case. If we compare the cases 
of one- and two-temperature plasma (left and right panels), we can observe that the regions where the difference between the SNB and SH models is noticeable 
appear qualitatively similar. Moreover, the features observed for the one-temperature regime (e.g. faster/slower damping in the case of the SNB model) are also present in the two-temperature regime. However, the difference between the SH and SNB models for wave periods is less pronounced in the two temperature 
regime. 

Additionally, the left panel of Fig.~\ref{fig:taus} shows the SH and SNB damping times exceed their values in one-temperature regime. This deviation can be 
explained in terms of different ion and electron temperatures. In this model, the thermal energy is transferred by electrons while the momentum is connected 
with ions. The only mechanism through which the wave can dissipate its energy is electron thermal conduction. However, the kinetic energy connected to the 
ions cannot be effectively dissipated because of the decoupling between the electron and the ion temperatures due to the $\tau_{ei}$ comparable to the wave 
period. It reduces the amount of energy the wave can dissipate through electron thermal conduction and, therefore, increases the damping time of slow waves in 
comparison to the one-temperature regime. Moreover, the presence of two temperatures complicates the onset of the isothermal regime: a uniform electron temperature established by strong thermal conduction can lose its uniformity when interacting with the non-uniform ion temperature, which is unaffected by thermal conduction. Thus, the onset 
of the isothermal regime is determined by both the electron thermal conduction and equilibration times. As can be seen from the left panel of 
Fig.~\ref{fig:taus}, SH damping times in both the one- and two-temperature regimes reach minima near 3\,MK, indicating that the equilibration between electron 
and ion temperatures occurs on a faster timescale than thermal conduction operates. In the SNB case, however, the damping time curve for the two-temperature regime
does not clearly reach a minimum unlike the one-temperature regime. This is attributed to the fact that electron 
thermal conduction and equilibration timescales are of similar orders.

\begin{figure*}
\includegraphics[width=\textwidth]{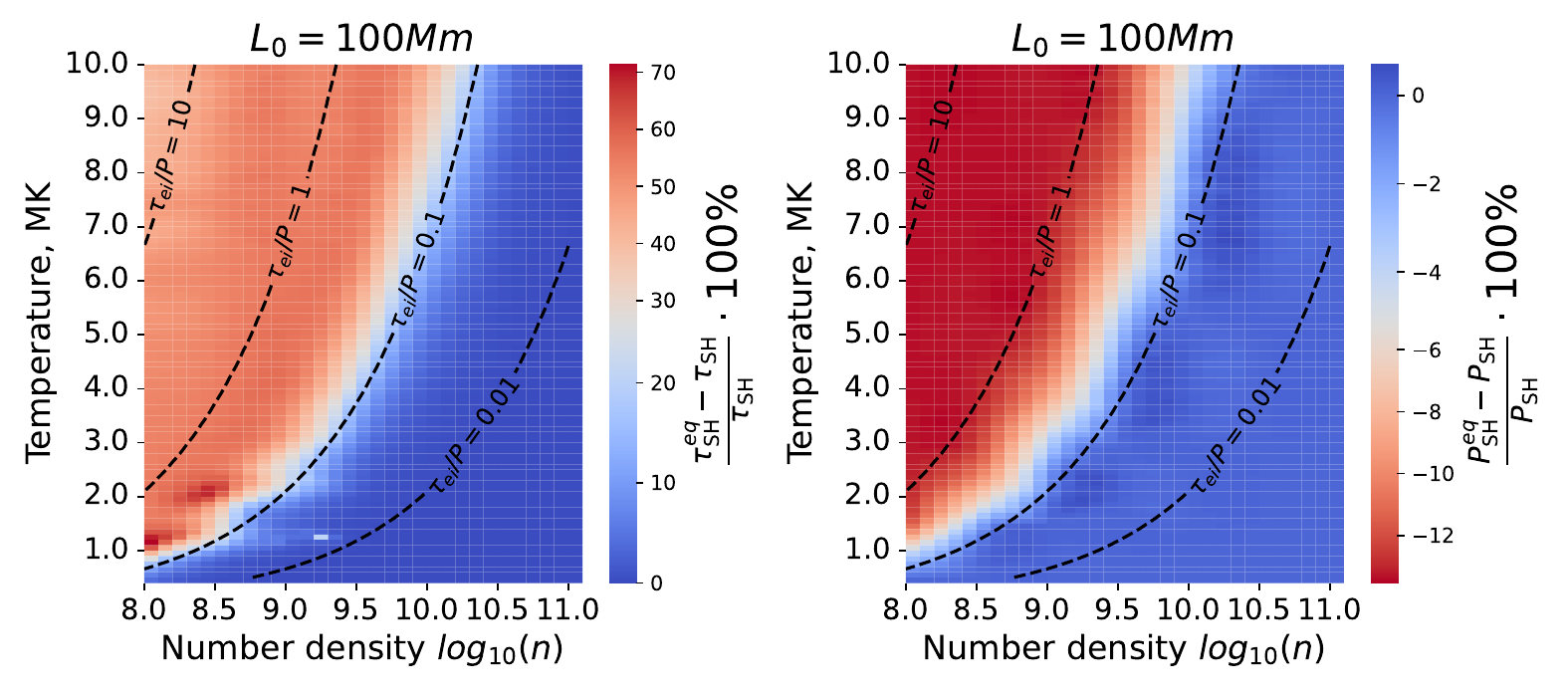}
    \caption{Relative difference between damping times (left panel) and periods (right panel) of slow waves obtained for the local thermal transport model in cases of one- and two-temperature plasma. The plot was obtained by using cubic interpolation of the results from $0.5\times0.5$ grid cell to $0.1\times0.1$ grid cell.}
    \label{fig:comparison_eq_SH}
\end{figure*}

The left panel of Fig.~\ref{fig:comparison_eq_SH}  plot shows the relative difference between damping time for the SH model in two- ($\tau^{eq}_{\mathrm{SH}}$) 
and one-temperature ($\tau_{\mathrm{SH}}$)  regimes, while the right panel demonstrates the difference between periods. It can be seen that finite equilibration 
results in higher damping times (up to 50\%) in comparison with the one-temperature regime. Moreover, the boundary of the region, where the deviation is 
noticeable, coincides with the $\tau_{ei}/P=0.1$ contour. It gives us an empirical condition $\tau_{ei}>0.1P$ when two-temperature effects are important for 
the dynamics of compressive waves. At the same time, the influence on periods is less pronounced and rarely exceeds 10\%. However, the same condition
$\tau_{ei}>0.1P$ is applicable in this case.

In summary, the addition of the finite equilibration delays  the onset of the transition of slow wave to isothermal regime due to the non-local thermal transport. Moreover, for plasma parameters corresponding to hot coronal loops, slow waves damp slower in the two-temperature regime.

\section{Conclusions}
\label{sec:conclusions}
In this study, we used the \Freyja~code to compare the influence of different thermal transport models on standing slow waves in 
one-temperature/two-temperature coronal loops in 1D. We considered a local thermal transport model (SH) \citep{Spitzer1953}, a flux-limiting thermal 
transport model (FL) \citep{OLSON2000619}, and the non-local thermal transport model (SNB) \citep{Schurtz2000, Brodrick2017} as thermal transport models.
Next, we simulated a fundamental standing slow-wave harmonic for all the thermal transport mechanisms and the plasma parameters listed in 
Table~\ref{tab:parameters}. 

For all simulations, we determined the wave period and the damping time by fitting the velocity perturbation at the center of the domain by an exponentially 
decaying sine function. We used the parameters determined to quantify how the different thermal transport models affected the 
observable parameters of standing slow waves. For both one- and two-temperature regimes, we observed the temperature-density regions where the non-local 
thermal transport model results in periods and damping times different from the case of the SH model. These regions are plotted in 
Fig.~\ref{fig:comparison_tau} and Fig.~\ref{fig:comparison_p} and coincide with the regions bounded by the Knudsen number contours.

We found that the damping time of standing slow waves for the SNB model can deviate from the case of the SH model. In the SNB case, standing waves can 
dissipate more rapidly or, in turn, more slowly with the border between these regimes around $K_n=0.01$. The reason for this damping time behaviour in the SNB
case is that the non-local thermal transport suppresses heat-fluxes, and it leads to the later appearance of the isothermal regime. As a result, the damping 
time minimum is shifted towards higher temperatures. Thus, for slow waves, smaller heat fluxes result in the thermal perturbation existing for a longer time 
that prevents the waves from being isothermal.

For the slow wave periods, the regions with significant differences between SNB and SH models were found. In all these regions, the SNB model showed the 
reduced periods and, therefore, higher phase speeds. This is explained, again, by the fact that the phase speed of slow waves decreases slower from sound 
speed to isothermal sound speed than in the case of local thermal transport because the non-local thermal transport prevents slow waves from switching to the 
isothermal regime. Thus, the periods are closer to their adiabatic values in the SNB case.

Moreover, we quantified the influence of the finite equilibration time on the slow wave dynamics. Our analysis showed that, in this case, damping times are up
to 50\% higher in comparison with the one-temperature regime. We explained it in terms of decoupling between ions and electrons. Also, we found an empirical 
condition $\tau_{ei}>0.1P$ showing when two-temperature effects should be taken into account. Using this condition, we concluded that the considered effect is
important for hot coronal loops.

Our parametric study showed that the non-local thermal transport may affect observable parameters of standing slow waves. We found that the non-local thermal transport effects 
are important for slow waves for a broad range of coronal parameters when $K_n>0.001$. For hot flaring loops parameters, it was shown that  the considered observable parameters are strongly affected by finite equilibration between ion and electron temperatures. The next perspective of this work is to study how the effect of the non-local thermal transport affects the polytropic index dependence on temperature detected in observations (see the references in Sec.~\ref{sec:intro}). Moreover, the use of the non-local thermal transport model for modelling plasma dynamics in solar flares is also of interest, since flares may have steeper temperature gradients in comparison to waves. 

{In our study, we considered a fully ionised hydrogen plasma.  In reality, the coronal composition is more complicated, and the presence of heavier elements is crucial for e.g. EUV emission and free-free radio emission. However, as hydrogen and helium form $\approx99.9\%$ and $\approx98\%$ of the Sun's number of atoms and mass, respectively, we do not expect that elements heavier than helium noticeably affect heat fluxes and mean plasma temperature. If helium is added, it results in a mean particle mass of about 0.6 of the hydrogen mass and effective ion charge number $Z_i\approx1.1$, which are close to the hydrogen values used in our study. Nevertheless, a detailed account for the solar composition is a promising future direction for the forward modelling of plasma processes with thermal conduction in both local and non-local regimes.}

\begin{acknowledgements}
      The work is funded by STFC Grant ST/X000915/1. DYK also acknowledges funding from the Latvian Council of Science Project No. lzp2022/1-0017.
\end{acknowledgements}

\bibliographystyle{aa}
\bibliography{refs}

\begin{thebibliography}{33}
\expandafter\ifx\csname natexlab\endcsname\relax\def\natexlab#1{#1}\fi

\bibitem[{Arber {et~al.}(2023)Arber, Goffrey, \& Ridgers}]{Arber2023}
Arber, T.~D., Goffrey, T., \& Ridgers, C. 2023, Frontiers in Astronomy and
  Space Sciences, 10

\bibitem[{Battaglia {et~al.}(2009)Battaglia, Fletcher, \& Benz}]{Battaglia2009}
Battaglia, M., Fletcher, L., \& Benz, A.~O. 2009, Astronomy \& Astrophysics,
  498, 891–900

\bibitem[{Brodrick {et~al.}(2017)Brodrick, Kingham, Marinak, Patel, Chankin,
  Omotani, Umansky, Del~Sorbo, Dudson, Parker, Kerbel, Sherlock, \&
  Ridgers}]{Brodrick2017}
Brodrick, J.~P., Kingham, R.~J., Marinak, M.~M., {et~al.} 2017, Physics of
  Plasmas, 24

\bibitem[{Caramana {et~al.}(1998{\natexlab{a}})Caramana, Burton, Shashkov, \&
  Whalen}]{caramana1998construction}
Caramana, E., Burton, D., Shashkov, M.~J., \& Whalen, P. 1998{\natexlab{a}},
  Journal of Computational Physics, 146, 227

\bibitem[{Caramana {et~al.}(1998{\natexlab{b}})Caramana, Shashkov, \&
  Whalen}]{caramana1998formulations}
Caramana, E.~J., Shashkov, M.~J., \& Whalen, P.~P. 1998{\natexlab{b}}, Journal
  of Computational Physics, 144, 70

\bibitem[{{De Moortel} \& {Hood}(2003)}]{2003A&A...408..755D}
{De Moortel}, I. \& {Hood}, A.~W. 2003, \aap, 408, 755

\bibitem[{Erdélyi \& Taroyan(2008)}]{Erdlyi2008}
Erdélyi, R. \& Taroyan, Y. 2008, Astronomy \& Astrophysics, 489, L49–L52

\bibitem[{Fleishman {et~al.}(2023)Fleishman, Nita, \& Motorina}]{Fleishman2023}
Fleishman, G.~D., Nita, G.~M., \& Motorina, G.~G. 2023, The Astrophysical
  Journal, 953, 174

\bibitem[{{Kolotkov}(2022)}]{Kolotkov2022}
{Kolotkov}, D.~Y. 2022, Frontiers in Astronomy and Space Sciences, 9, 402

\bibitem[{{Kolotkov} {et~al.}(2020){Kolotkov}, {Duckenfield}, \&
  {Nakariakov}}]{2020A&A...644A..33K}
{Kolotkov}, D.~Y., {Duckenfield}, T.~J., \& {Nakariakov}, V.~M. 2020, \aap,
  644, A33

\bibitem[{Kolotkov {et~al.}(2019)Kolotkov, Nakariakov, \&
  Zavershinskii}]{Kolotkov2019}
Kolotkov, D.~Y., Nakariakov, V.~M., \& Zavershinskii, D.~I. 2019, Astronomy \&
  Astrophysics, 628, A133

\bibitem[{Mariska(2005)}]{Mariska2005}
Mariska, J.~T. 2005, The Astrophysical Journal, 620, L67–L70

\bibitem[{Mariska(2006)}]{Mariska2006}
Mariska, J.~T. 2006, The Astrophysical Journal, 639, 484–494

\bibitem[{Mariska {et~al.}(2008)Mariska, Warren, Williams, \&
  Watanabe}]{Mariska2008}
Mariska, J.~T., Warren, H.~P., Williams, D.~R., \& Watanabe, T. 2008, The
  Astrophysical Journal, 681, L41–L44

\bibitem[{Nakariakov {et~al.}(2019)Nakariakov, Kosak, Kolotkov, Anfinogentov,
  Kumar, \& Moon}]{Nakariakov2019}
Nakariakov, V.~M., Kosak, M.~K., Kolotkov, D.~Y., {et~al.} 2019, The
  Astrophysical Journal Letters, 874, L1

\bibitem[{Nisticò {et~al.}(2017)Nisticò, Polito, Nakariakov, \&
  Del~Zanna}]{Nistic2017}
Nisticò, G., Polito, V., Nakariakov, V.~M., \& Del~Zanna, G. 2017, Astronomy
  \& Astrophysics, 600, A37

\bibitem[{Ofman \& Wang(2022)}]{Ofman2022}
Ofman, L. \& Wang, T. 2022, The Astrophysical Journal, 926, 64

\bibitem[{Olson {et~al.}(2000)Olson, Auer, \& Hall}]{OLSON2000619}
Olson, G.~L., Auer, L.~H., \& Hall, M.~L. 2000, Journal of Quantitative
  Spectroscopy and Radiative Transfer, 64, 619

\bibitem[{Paddock {et~al.}(2023)Paddock, Li, Kim, Lee, Martin, Ruskov, Hughes,
  Rose, Murphy, Scott, {et~al.}}]{paddock2023energy}
Paddock, R., Li, T., Kim, E., {et~al.} 2023, Plasma Physics and Controlled
  Fusion, 66, 025005

\bibitem[{Pant {et~al.}(2017)Pant, Tiwari, Yuan, \& Banerjee}]{Pant2017}
Pant, V., Tiwari, A., Yuan, D., \& Banerjee, D. 2017, The Astrophysical Journal
  Letters, 847, L5

\bibitem[{Prasad {et~al.}(2018)Prasad, Raes, Van~Doorsselaere, Magyar, \&
  Jess}]{Prasad2018}
Prasad, S.~K., Raes, J.~O., Van~Doorsselaere, T., Magyar, N., \& Jess, D.~B.
  2018, The Astrophysical Journal, 868, 149

\bibitem[{{Reale} {et~al.}(2019){Reale}, {Testa}, {Petralia}, \&
  {Kolotkov}}]{2019ApJ...884..131R}
{Reale}, F., {Testa}, P., {Petralia}, A., \& {Kolotkov}, D.~Y. 2019, \apj, 884,
  131

\bibitem[{Schurtz {et~al.}(2000)Schurtz, Nicolaï, \& Busquet}]{Schurtz2000}
Schurtz, G.~P., Nicolaï, P.~D., \& Busquet, M. 2000, Physics of Plasmas, 7,
  4238–4249

\bibitem[{Spitzer \& H\"{a}rm(1953)}]{Spitzer1953}
Spitzer, L. \& H\"{a}rm, R. 1953, Physical Review, 89, 977–981

\bibitem[{Van~Doorsselaere {et~al.}(2011)Van~Doorsselaere, Wardle, Del~Zanna,
  Jansari, Verwichte, \& Nakariakov}]{VanDoorsselaere2011}
Van~Doorsselaere, T., Wardle, N., Del~Zanna, G., {et~al.} 2011, The
  Astrophysical Journal, 727, L32

\bibitem[{Wang \& Ofman(2019)}]{Wang2019}
Wang, T. \& Ofman, L. 2019, The Astrophysical Journal, 886, 2

\bibitem[{Wang {et~al.}(2015)Wang, Ofman, Sun, Provornikova, \&
  Davila}]{Wang2015}
Wang, T., Ofman, L., Sun, X., Provornikova, E., \& Davila, J.~M. 2015, The
  Astrophysical Journal, 811, L13

\bibitem[{Wang {et~al.}(2018)Wang, Ofman, Sun, Solanki, \& Davila}]{Wang2018}
Wang, T., Ofman, L., Sun, X., Solanki, S.~K., \& Davila, J.~M. 2018, The
  Astrophysical Journal, 860, 107

\bibitem[{{Wang} {et~al.}(2021){Wang}, {Ofman}, {Yuan}, {Reale}, {Kolotkov}, \&
  {Srivastava}}]{2021SSRv..217...34W}
{Wang}, T., {Ofman}, L., {Yuan}, D., {et~al.} 2021, \ssr, 217, 34

\bibitem[{Wang {et~al.}(2002)Wang, Solanki, Curdt, Innes, \&
  Dammasch}]{Wang2002}
Wang, T., Solanki, S.~K., Curdt, W., Innes, D.~E., \& Dammasch, I.~E. 2002, The
  Astrophysical Journal, 574, L101–L104

\bibitem[{Wang {et~al.}(2003)Wang, Solanki, Curdt, Innes, Dammasch, \&
  Kliem}]{Wang2003}
Wang, T.~J., Solanki, S.~K., Curdt, W., {et~al.} 2003, Astronomy \&
  Astrophysics, 406, 1105–1121

\bibitem[{West {et~al.}(2008)West, Bradshaw, \& Cargill}]{West2008}
West, M.~J., Bradshaw, S.~J., \& Cargill, P.~J. 2008, Solar Physics, 252,
  89–100

\bibitem[{{West} {et~al.}(2004){West}, {Cargill}, \&
  {Bradshaw}}]{West2004ESASP.575..573W}
{West}, M.~J., {Cargill}, P.~J., \& {Bradshaw}, S.~J. 2004, in ESA Special
  Publication, Vol. 575, SOHO 15 Coronal Heating, ed. R.~W. {Walsh},
  J.~{Ireland}, D.~{Danesy}, \& B.~{Fleck}, 573

\end{thebibliography}

\end{document}